\documentclass[prl,twocolumn, superscriptaddress]{revtex4}
\usepackage{graphicx,amsmath,amssymb,amsxtra}
\renewcommand{\narrowtext}{\begin{multicols}{2} \global\columnwidth20.5pc}

\def\be{\begin{eqnarray}}
\def\ee{\end{eqnarray}}

\newcommand{\Eq}[1]{Eq.~(\ref{#1})}
\newcommand{\Fig}[1]{Fig.~(\ref{#1})}

\newcommand{\ket}[1]{{| #1\rangle }}

\begin{document}
%\draft

\title{S-duality constraints on 1D patterns associated with fractional quantum Hall states}

\author{Alexander Seidel}
\affiliation{Department of Physics and Center for Materials Innovation, Washington University, St. Louis, MO 63136, USA}

\date{\today}

\begin{abstract}
Using the modular invariance of the torus, constraints on the 1D patterns are derived
that are associated with various fractional quantum Hall ground states, e.g. through the
thin torus limit. In the simplest case, these constraints enforce the well known odd-denominator
rule, which is seen to be a necessary property of all 1D patterns associated to quantum Hall states with
minimum torus degeneracy. However, the same constraints also have implications for the
 non-Abelian states possible within this framework. In simple cases, including the $\nu\!=\!1$
 Moore-Read state and the $\nu\!=\!3/2$ level 3 Read-Rezayi state, the filling factor and the torus
 degeneracy uniquely specify the possible patterns, and thus all physical properties that are
 encoded in them. It is also shown that some states, such as the "strong p-wave pairing state", cannot in principle be described through  patterns.
\end{abstract}

\maketitle
{\em Introduction.}
The study
 of fractional quantum Hall (FQH) liquids
has  been among the most intriguing problems 
in condensed matter physics during the past few decades, in both theory and experiment.
On the theoretical side, the construction of variational many-body
wave functions has traditionally played a pivotal role\cite{laughlin1}.
In principle, the possible variational constructions are limitless.
A systematic classification of FQH phases therefore requires
additional constraints, such as simplicity in a composite fermion picture\cite{jain}.
Another program to implement such constraints
is to require that the trial wave functions can be obtained as conformal blocks
in certain conformal field theories\cite{mooreread}. 
The problem is then relegated to identifying all conformal field theories
leading to permissible wave functions. 
On the other hand, it has recently become appreciated that a large
class of trial wave functions can be characterized
by simple sequences of integers, either through the thin torus limit 
and adiabatic continuity\cite{seidel1,seidel2, seidelyang, karlhede1,karlhede2,karlhede3},
or through Jack polynomials\cite{haldanebernevig}.
The patterns of integers associated with viable quantum Hall
states are in turn subject to a number of consistency requirements,
such as rotational invariance of the associated Jack polynomials,
or constraints on the associated "patterns of zeros" studied in
Ref. \onlinecite{wenwang}.
A complete set of consistency requirements 
is desirable in order to understand the possible quantum numbers
of all quantum Hall phases that are accessible within this framework.
In this paper, it will be shown that 
the one-dimensional (1D) patterns associated with the ground
state sectors of a quantum Hall phase are highly constrained
by modular invariance on the torus. 
In the simplest case, the implication of these constraints is the well-known
"odd-denominator-rule", which is found to be required
within this framework
for all
quantum Hall states that have the "minimum torus degeneracy". Such states
are necessarily Abelian, and include states in the Haldane-Halperin hierarchy.
Furthermore, in some other cases of interest, it is found that 
the filling factor and the torus degeneracy already completely
determine the associated set of 1D patterns. These patterns, in turn, can be shown
to have direct implications for the statistics of quasiparticles, using the method
of Ref. \onlinecite{seidel_pfaff}. This is in particular true for the Moore-Read
state\cite{mooreread}, where the statistics are fully determined
modulo a certain Abelian phase\cite{seidel_pfaff}. Similar statement apply\cite{JFU} to the $k=3$ Read-Rezayi state\cite{readrezayi}.
It is thus found that within this framework, the filling factor and the torus degeneracy alone
may greatly constrain the low energy physics in some cases.
Furthermore, it will also be shown
that certain states do not admit a description in terms of
1D patterns. This includes the well known "strong p-wave pairing" state
at filling factor 1/2. While this does not directly imply that states of this kind are 
not
physical, 
the possibility to sharply distinguish between states that 
do admit
a 1D representation and  states that do not may
hint at  qualitative difference between these two categories.

\indent{\em 1D patterns and S-duality.}
The bulk properties of fractional quantum Hall phases are
expected to be independent of the topology of the two-dimensional (2D) space they 
live in. In the present work the topology of choice
 will be the torus,
 which is identified with
a rectangular
2D domain of dimensions $L_x$ and $L_y$ subject to
periodic (magnetic) boundary conditions. 
%I will set the
%magnetic length equal to $1$, such that $L_xL_y=2\pi L$,
%where L is the number of magnetic flux quanta 
%through the surface, or the number of orbitals in the
%lowest Landau level (LLL).
It will be assumed that the state under consideration has a 
well defined thin torus limit in which simple 1D patterns
given by Landau level (LL) orbital occupancies emerge.
For definiteness it may be assumed that adiabatic continuity holds
\cite{seidel1, seidel2,seidelyang, karlhede1, karlhede2, karlhede3} between the
2D limit of a large torus and the 1D thin torus limit.
However, this is not essential in the following arguments,
as long as a set of "dominance patterns" can be obtained through a
formal 1D limit, which carry the correct quantum numbers of the ground states under translations.
Furthermore, such a set of patterns is also manifest in 
the Jack polynomial construction\cite{haldanebernevig}, even though
the latter is not available on the torus.
As an example, let us consider the patterns that arise
in the thin torus limit for the threefold degenerate ground state
of the $\nu=1$ Pfaffian, i.e.  $20202020\dotsc$,
$02020202\dotsc$ and $11111111\dotsc$ \cite{seidel2}.
In the limit  $L_y\rightarrow 0$, these labels
describe the definite occupancy numbers
of the limiting ground states in a 
certain basis of LL orbitals $\varphi_n$. 
This basis is taken to consist of
orbitals that are localized in $x$ and wrap around the torus
in the $y$ direction. 
On the other hand, the roles of $x$ and $y$ 
can be exchanged through the modular S-transformation,
which leaves the physics invariant.
Hence the same patterns can be obtained in the "dual" 
limit $L_x\rightarrow 0$, where they appear as occupation
numbers in a dual LL basis $\bar\varphi_n$. 
The $\bar\varphi_n$  can be thought
of as a ``rotated'' version of the $\varphi_n$,
and  are related to the $\varphi_n$ via
Fourier transform % in $n$
(see, e.g., Ref.\onlinecite{seidel3}).

A rather stringent constraint on legitimate thin torus
patterns can be obtained by exploring the
consequences of modular S-invariance
on the representation of the magnetic translation group
formed by the ground states. 
The magnetic translation group
is generated by operators $T_x$ and $T_y$
that act on single particle LL orbitals via
\begin{subequations} \label{t}
\begin{align}
\label{Tphi} 
T_x \varphi_n(z) & = \varphi_{n+1}(z) &
T_y \varphi_n(z) & = e^{-\frac{2 \pi i}{L} n}\varphi_n(z)
 \\ 
\label{Tvphi} 
 T_x \overline{\varphi}_n(z) & = e^{\frac{2 \pi i}{L} n} \overline{\varphi}_n(z) &
T_y \overline{\varphi}_n(z) & = \overline{\varphi}_{n+1}(z),
\end{align}
\end{subequations}
cf., e.g., Ref. \onlinecite{seidel3}. Here, $L$ is the total number of LL orbitals.
The torus ground states of  a given quantum Hall phase
form a representation of the magnetic translation group.
This representation cannot depend on the aspect ratio of the torus.
If a simple thin torus limit exists in the sense described above,
it allows one to immediately infer the matrices $R(T_x)$, $R(T_y)$
of this representation.
In the limit $L_y\rightarrow 0$, where the patterns
extend along the $x$ direction and
correspond to simple product states in the $\varphi_n$ basis, $\eqref{Tphi}$
implies that such product states are eigenstates of $T_y$ with eigenvalue
$\exp(-2\pi i/ L \, \sum_j n_j)$. Here, $n_j$ is the orbital index of the $j$-th particle
in the pattern. Likewise, $T_x$ performs a right-shift of the pattern.

Physically, modular S-invariance on the torus 
 is the statement that the $x$ and $y$ coordinates of the system play
 interchangeable roles. This means that in the opposite thin torus limit,
$L_x\rightarrow 0$, the {\em same} thin torus patterns must appear.
These patterns now extend along $y$ and correspond to simple 
product states in the $\overline{\varphi}_n$ basis.
In general, however, a ground state that evolves into a given
1D pattern in one thin torus limit will evolve into
a superposition of such patterns in the opposite
thin torus limit, and vice versa. As a result, the representation
matrices obtained from the product ground states (patterns)
in the two mutually "dual" thin torus limits are unitarily equivalent, 
but not identical. \Eq{t} immediately implies the following relations
when passing from the $L_y\rightarrow 0$ limit to the dual
limit $L_x\rightarrow 0$:
%If we use these states to calculate the representation matrices of $T_x$
%and $T_y$, these will be related to the original matrices,
%obtained in the limit $L_y\rightarrow 0$, by the ``rotation'':
\begin{equation}
  \label{rotation}
   %\begin{split}
    R(T_x) =\bar R(T_y)\;,\qquad
    R(T_y) = \bar R(T_x)^\dagger\;.
   %\end{split}
\end{equation}
%where $R(T_{x,y})$ denotes the matrix representing $T_{x,y}$.
In the above, $R(T_{x,y})$ refers to the matrices 
describing $T_x$ and $T_y$ in the basis of product ground states
emerging in the $L_y\rightarrow 0$ limit.
%These are product states in the $\varphi_n$ LL basis.
Let us label these states by $\ket{\tau}$, where $\tau$
denotes the associated simple pattern, e.g. $\tau=2020\dotsc, 1111\dotsc, 0202\dotsc$
for the $\nu=1$ Pfaffian.  $\bar R(T_{x,y})$ are the matrices describing
 the transformation properties of the
dual product states $\overline{\ket{\tau}}=S\ket{\tau}$, where
$S$ is the unitary transformation that takes the orbital $\varphi_n$ into
$\overline\varphi_n$. Eqs. \eqref{rotation} then immediately follow
from Eqs. \eqref{t}. Apparently, the state $\ket{\tau}$ in general
has different transformation properties from its dual version $\overline{\ket{\tau}}$.
Hence these states correspond to opposite thin torus limits of 
{\em different} degenerate ground states.
On the other hand, so long as the ground state patterns
describe the translational properties of the ground states for any aspect ratio of the torus
(as implied, e.g., by adiabatic continuity),
the matrices $R$ and $\bar R$ must form the same representation.
They must therefore be related by a unitary transformation, $\bar R(T_{x,y})=U^\dagger R(T_{x,y})U$.
\Eq{rotation} then becomes:
%\abovedisplayshortskip=6pt minus 4pt
%\belowdisplayshortskip=6pt minus 4pt
%\abovedisplayskip=6pt minus 4pt
%\belowdisplayskip=6pt minus 4pt
\begin{equation}
  \label{duality}
   %\begin{split}
    R(T_x) = U^\dagger R(T_y) U\;,\qquad
    R(T_y) = U^\dagger R(T_x)^\dagger U\;.
   %\end{split}
\end{equation}
The above says that the representation of the magnetic translation 
group implied by the patterns must be "selfdual", i.e. the matrices
associated with $T_x$ and $T_y$ are interchangeable in the precise sense
of \Eq{duality}. From given ground state patterns, it is always easy to work
out these matrices from \eqref{t} as described above. \Eq{duality}
then poses severe constraints on these patterns.

{\em Odd denominator rule.}
%\abovedisplayshortskip=6pt minus 4pt
%\belowdisplayshortskip=6pt minus 4pt
%\abovedisplayskip=6pt minus 4pt
%\belowdisplayskip=6pt minus 4pt
%\begin{equation}
 % \label{Ty}
 % T_y=\exp(\frac{2\pi i}{N_\phi} \sum_i n_i)
%\end{equation}
A non-trivial torus degeneracy is a hallmark of
topologically ordered systems. It is well understood that quite generally, if the
system is characterized by a filling factor $\nu=p/q$, with
$p$ and $q$ co-prime, its minimum torus degeneracy
is $q$ \cite{oshikawa2, hastings}.  
This lower bound is typically exceeded in time-reversal invariant
topologically ordered systems. 
It is, however, satisfied for the simplest fractional 
quantum Hall states, such as those
in the Abelian hierarchy. Hierarchy states are also known 
for their compliance with the "odd denominator rule",
according to which $q$ must be odd. This has been understood in
various ways \cite{su,tao}. Here it will be shown
that for all states that can be represented through
periodic 1D patterns as discussed, the odd denominator
rule is a direct consequence of the minimum torus degeneracy,
together with the requirement \eqref{duality}.

\begin{figure}[t]
\begin{center}
\includegraphics[width=5cm]{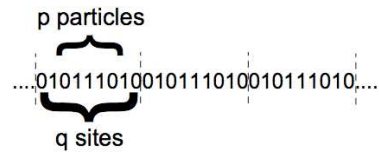}
\end{center}
\caption{\label{figure} A generic periodic 1D pattern as it might appear in the thin torus limit
of a fractional quantum Hall state, for a given topological sector. In addition, the individual
orbitals may carry a pseudospin label, as in Ref. \onlinecite{seidelyang}. The complete
set of patterns for all topological sectors are subject to the duality constraint \eqref{duality}.}
\end{figure}

Suppose, now, an incompressible quantum Hall state
can be represented by a 1D pattern with a unit cell
containing $p$ particles and $q$ orbitals, Fig. \ref{figure}, such that the LL
filling factor is $\nu=p/q$.
By translational symmetry, there must be at least $q$ ground states
on the torus, since evidently $q$ translations are required to transform
the state back to itself. If we further assume that the state has the minimum
torus degeneracy, it follows that {\em all} ground states are related by translation
{\em and} that $p$ and $q$ are co-prime. Instead, however, I will start
from the weaker assumption that all ground states are related
by translation.  Using \Eq{duality}, this is already sufficient to show
 that $p$ and $q$ are co-prime, and hence the state has
the minimum torus degeneracy. To see this,  let the state associated
with the pattern be denoted by $\ket{\tau}$. The states $T_x^j\ket{\tau}$, $j=0...q-1$
then represent a complete set of torus ground states. From these, we can easily form
eigenstates of $T_x$ with $q$ distinct eigenvalues.
In contrast, $T_y$ is found to have $q$ distinct eigenvalues only if $p$ and $q$
are co-prime. Note that the states $T_x^j\ket{\tau}$ are already eigenvalues
of $T_y$ with eigenvalues that can be read off directly from the associated patterns as
described above. Each application of $T_x$ changes the $T_y$ eigenvalue by a factor
$\exp(-2\pi i \nu)$. It follows from this that if $\nu=p/q=p'/q'$ where $p'$ and $q'$ are 
co-prime, the state $\ket{\tau}$ and its translated counterparts have exactly 
$q'$ distinct $T_y$ eigenvalues. The $S$-duality requirement \Eq{duality}
then implies that $q=q'$ since $T_x$ and $T_y$ must in particular have
the same spectrum within the ground state space. Hence also $p=p'$.
Thus one finds that whenever all torus ground states of a specific quantum Hall
phase are related by translation,
any permissible 1D pattern associated with this phase must satisfy that the size $q$ of
its unit cell and the number $p$ of particles contained therein are co-prime.
To proceed, let us further assume that the total number $N$ of particles
in the state is even. This can be done without any loss of generality, since
when some incompressible quantum Hall fluid exists for odd particle number on the torus, 
it also exists for even particle number by means of doubling the system size.
The operators $T_x^q$ and $T_y^q$ are constant (proportional to the identity)
within the space spanned by the $q$  ground states, since $\ket{\tau}$ is an
eigenstate of both, and both operators commute with $T_x$.
By the duality constraint \Eq{duality}, both operators must therefore be equal.
By acting on $\ket{\tau}$, one easily finds that $T_x^q=(-1)^{fp}$, where $f=1$ for
fermions and $f=0$ for bosons. This follows since, with standard phase conventions,
each fermion that is translated from the $L$-th orbital  to the $1$st one will give rise to
a negative sign, as is must be commuted through $(N-1)$ occupied fermion states.
This happens exactly $p$-times when the operator $T_x$ is applied $q$-times to the product state
associated with $\ket{\tau}$  (see \Fig{figure}). On the other hand,
$T_y^q=(-1)^{pq}$. Again, we evaluate this by acting on $\ket{\tau}$.
According to the prescription below \Eq{t},
\begin{equation}\label{tyq}
T_y^q\ket{\tau}=e^{-\frac{2\pi iq}{L}\sum_{j=1}^Ln_j}\ket{\tau}=e^{-\frac{2\pi i}{M}\sum_{k=1}^M(u+(k-1)qp)}\ket{\tau}
\end{equation}
where $M$ is the number of unit cells in the pattern such that $L=Mq$, $N=Mp$.
The integer $u=\sum_{j=1}^p n_j$ equals the contribution of the first
unit cell to the sum such that the $k$th unit cell contributes
$u+(k-1)pq$. Since the $u$-term drops out modulo $2\pi i$,
the exponent in \eqref{tyq} reads $-i\pi (M-1) pq=i \pi pq  \mod 2\pi i$,
since $Mp=N$ was assumed to be even. Hence $T_y^q=(-1)^{pq}$.

One thus finds that $(-1)^{pq}=(-1)^{fp}$ for any
quantum Hall state that can be represented
through 1D patterns, whenever all torus ground states
are related by translation. This implies that such
states satisfy the odd denominator rule: If $q$ were
even, $p$ would have to be odd, and the relation
would be violated for fermions. It likewise
follows that for bosons, out of $p$ and
$q$ exactly one needs to be even\cite{su,tao}.
Within this framework, the odd denominator rule
(and its bosonic counterpart) has thus been shown
to be a characteristic property of all states with minimum
torus degeneracy.
It should be noted that for all states in the
Abelian hierarchy, bosonic and fermionic, 1D patterns have been worked out in Ref. \onlinecite{karlhede5}.
It is pleasing to see that with the above considerations,
 it can be understood
from the patterns themselves that some are
legitimate for bosonic states only, while some only 
qualify for fermionic states.

{\em The strong p-wave pairing state.}
Quantum Hall states that satisfy the minimum torus degeneracy
are necessarily Abelian. Within the framework of 1D patterns,
this follows from the fact that if all patterns are related
by translation, domain walls associated with elementary quasiparticle type excitations
always generate the same fixed shift between subsequent ground state patterns.
In this case, the degeneracy of topological sectors also remains fixed (cf., e.g., \cite{seidel3}),
and does not grow exponentially with quasiparticle number as required for non-Abelian states.
Conversely, however, Abelian states need not satisfy the minimum torus degeneracy,
and can thus violate the odd denominator rule. 
Examples are found among the Halperin
bilayer states, whose thin torus patterns have been given
in \cite{seidelyang}. While the patterns of such
states are not all related by translation, they do all have unit cells
of the same size. This must be true in order for the states to be Abelian.
A variation in unit cell size between different ground state patterns
will always introduce a combinatorial degree of freedom when domain walls
between different patterns are formed, %. Namely, by varying the length of the
%patterns with the shorter unit cell, variable relative shifts between the patterns
%with the longer unit cell are created, as demonstrated
as becomes clear, e.g., by considering the Pfaffian
case \cite{seidel2,karlhede3,seidel_pfaff}. In the interpretation
of Ref. \onlinecite{ardonne4}, this always leads to non-trivial
fusion rules, implying a non-Abelian state.
 On the other hand, different patterns of equal unit cell size may 
 or may not do this. (They do so for the level 3 Read-Rezayi state,
 with patterns given below.)
 % also 
 %realize this, as is the case e.g. for the level 3 Read-Rezayi state\cite{readrezayi}
 %(see below for patterns).
With this in mind, it is interesting to ask whether
 an Abelian single-component state of fermions at $\nu=1/2$ with an
eight-fold torus degeneracy can be consistent with the framework
described here.
These are the quantum numbers relevant to the
Abelian state now known as the ``strong p-wave pairing'' state\cite{readgreen}, 
which was originally discussed in Ref. \onlinecite{greiter}
as a candidate for the plateau at $\nu=5/2$ \cite{nu52}.
Here we can easily rule out that this state fits
into the 1D formalism.
The elementary unit cell of the corresponding 1D pattern
could not have size $8$, for then all eight ground states must
be related
by translation. This can be ruled out, since the state must then be subject 
to the odd denominator rule as shown above.
Alternatively, an Abelian state at $\nu=1/2$
could correspond to
1D patterns formed from
two different unit cells
of size $4$, or four different unit cells of size $2$,
However, this is not possible either, since for fermions
at $\nu=1/2$ there is only one type of elementary unit cell,
modulo translations, of size $4$ or size $2$. 
These unit cells are $1100$ and $10$, respectively.
This rules out that any Abelian state 
with the quantum numbers of the
strong pairing state
can be described in the language of 1D patterns.
In fact, we can also rule out a non-Abelian state with these
quantum numbers. Three different ground state patterns
of unit cell sizes $4$,$2$,$2$ can be ruled out as in the above.
Two ground state patterns of unit cell sizes $6$ and $2$ are
again found to violate the S-duality constraint \eqref{duality}:
As shown above, at $\nu=1/2$ any given pattern, including its translated versions,
can only
account for 2 different $T_y$ eigenvalues. However, there must be at least
$6$ different $T_x$ eigenvalues if a pattern has unit cell size $6$.

Needless to say, 
the incompatibility of the strong pairing state with a description
in terms of 1D patterns
does not necessarily rule out the
viability of such a state.
It may, however, imply  
that this state is of a qualitatively different nature
compared to other contenders that 
do allow a 1D labeling, such as the $\nu=1/2$
Pfaffian.
In this regard it is worth noting that %, to the best of my knowledge,
so far the strong pairing state seems to have been quite elusive to exact diagonalization
studies.

{\em Non-Abelian and other states.}
I finally remark that the considerations
made above allow one, in simple enough cases,
to positively identify the possible quantum Hall sates
allowed within the 1D formalism, based on the
filling factor and the torus degeneracy alone.
Indeed, within this framework, these two data
may specify the underlying physics quite uniquely.
As an example, I will analyze the question
of how many possible bosonic quantum Hall states
may exist at filling $\nu=1$ with a threefold torus
degeneracy, which fit into the 1D framework.
This is easily answered. The pattern $300300\dotsc$
can be ruled out, since it already accounts for
a $3$-fold degeneracy. Hence all ground states would
be related by translation. However, the state violates
the bosonic analogue of the odd denominator rule,
and so would then violate \Eq{duality}, as shown above.
Patterns with unit cell sizes $2$ and $1$ are unique
at $\nu=1$, and must then constitute the correct ground state
patterns. These are $2020\dotsc$ and $1111\dotsc$, respectively,
the patterns associated with the $\nu=1$ Pfaffian.
These satisfy S-duality, as already hinted at in Ref.\onlinecite{seidel2}.
Furthermore, it has been shown how these patterns 
essentially encode the statistics of the state\cite{seidel_pfaff},
modulo a certain Abelian phase.
The filling factor
$\nu=1$
and the torus degeneracy $3$ thus specify the physics
quite uniquely within the framework discussed here.
Similar statements
can be made about Laughlin states. 
Moreover,
the same constraints also fix the patterns of the level $3$ Read-Rezayi state
at $\nu=3/2$ ($4$-fold degenerate). Here, a single pattern of unit cell size $4$
is ruled out: Such a unit cell must contain $6$ particles, 
which is not co-prime with $4$, 
in violation of the rules established above.
Two patterns of unit cell size $2$ is the only
other possibility that can account for these quantum
numbers. This uniquely determines the patterns to be $3030\dotsc$
and $2121\dots$, which are just the patterns that have been associated to this state
in the literature\cite{haldanebernevig, ardonne4}.
Last, let us inquire about a state at fillig factor $\nu=2/3$,
with torus degeneracy 6. These are the quantum numbers of the bosonic
gaffnian\cite{gaffnian}, which, unlike the other states discussed so
far, has been proposed to be critical. Irrespective of 
its physical nature, the associated patterns\cite{haldanebernevig,ardonne5} again turn out to be
unique based on these quantum numbers.
% the
%unique patterns that are consistent with these quantum numbers.
At $\nu=2/3$, possible unit cell sizes must be multiples of $3$.
A single pattern of unit cell size $6$ can be ruled out as in the
case of the Read-Rezayi state. There must then be 2 patterns
of unit cell size $3$. These are again unique, modulo translation:
$200\dots$ and $101\dotsc$. It remains to be seen if even in this -- presumably
critical -- case, the method of Ref.\onlinecite{seidel_pfaff} can be used to 
ascribe well defined statistics to this state.

\begin{acknowledgments}
I would like to thank A. Karlhede, E. Bergholtz, and D.H. Lee for stimulating discussions.
This work was supported  by the National Science Foundation under NSF Grant No. DMR-0907793.
\end{acknowledgments}

%\bibliography{odden}

\end{document}